# A Multimodal Assistive System for Product Localization and Retrieval for People who are Blind or have Low Vision

Ligao Ruan[a], Giles Hamilton-Fletcher[c,d], Mahya Beheshti[a,c], Todd E Hudson[c], Maurizio Porfiri[a,b,e,f] and John-Ross Rizzo*[a,b,d,e]

[a]*Department of Mechanical and Aerospace Engineering, NYU Tandon School of Engineering, Brooklyn, New York, USA;* [b]*Department of Biomedical Engineering, NYU Tandon School of Engineering, Brooklyn, New York, USA;* [c]*Department of Rehabilitation Medicine, NYU Grossman School of Medicine, New York, New York, USA;* [d]*Department of Radiology, NYU Grossman School of Medicine, New York, New York, USA;* [e]*Center for Urban Science and Progress, NYU Tandon School of Engineering, Brooklyn, New York, USA;* [f]*Department of Civil and Urban Engineering, NYU Tandon School of Engineering, Brooklyn, New York, USA*

John-Ross Rizzo, MD, Rehabilitation Medicine, Vice Chair for Innovation & Equity, Ilse Melamid Endowed A/Prof of Rehab Medicine, Associate Professor of Neurology, Mechanical & Aerospace Engineering, and Biomedical Engineering

Email: JohnRoss.Rizzo@nyulangone.org

# A Multimodal Assistive System for Product Localization and Retrieval for People who are Blind or have Low Vision

Abstract: Shopping is a routine activity for sighted individuals, yet for people who are blind or have low vision (pBLV), locating and retrieving products in physical environments remains a challenge. This paper presents a multimodal wearable assistive system that integrates object detection with vision-language models to support independent product or item retrieval, with the goal of enhancing users' autonomy and sense of agency. The system operates through three phases: product search, which identifies target products using YOLO-World detection combined with embedding similarity and color histogram matching; product navigation, which provides spatialized sonification and VLM-generated verbal descriptions to guide users toward the target; and product correction, which verifies whether the user has reached the correct product and provides corrective feedback when necessary. Technical evaluation demonstrated promising performance across all modules, with product detection achieving near-perfect accuracy at close range and high accuracy when facing shelves within 1.5 m. VLM-based navigation achieved up to 94.4% accuracy, and correction accuracy exceeded 86% under optimal model configurations. These results demonstrate the system's potential to address the last-meter problem in assistive shopping. Future work will focus on user studies with pBLV participants and integration with multi-scale navigation ecosystems.

**Introduction**

Product retrieval is a fundamental component of independent shopping across diverse retail environments, including grocery stores, pharmacies, department stores, and clothing retailers [1]. Standing before product displays, shoppers must distinguish dozens of similar-looking packages, assess their spatial layout, and coordinate reaching and grasping motions to retrieve the target product. pBLV often cannot access, or can only partially access, the packaging colors, textures, or printed labels that typically guide these micro-decisions [2]. Consequently, they must either spend considerable time relying on residual vision to identify products or seek assistance from store staff or other customers [3]; however, research has shown that both approaches can be inefficient and would degrade the shopping experience [4][5]. When no assistance is available, they often purchase the wrong items or abandon their shopping altogether, resulting in diminished autonomy and reduced quality of life [6][7].

Over the past decade, various assistive technology solutions have emerged to support activities of daily living for pBLV [8–13]. Advancements in artificial intelligence (AI), particularly in computer-vision models [14] and vision-language models (VLMs) [15], have enabled assistive systems to extract and interpret visual information from input. These AI-based systems can receive camera feeds, process frames in real time, and generate haptic or audio feedback for users [16][17][18]. Such capacities have been applied across multiple scenarios, including wayfinding [19][20], obstacle avoidance [21][22], and commuting assistance [23], enhancing the independent living capabilities for pBLV.

However, most existing assistive systems still fall short of providing comprehensive support for pBLV to locate and retrieve products in stores. For example, the system proposed by [24] employed an object detection model to identify products and utilized GPS for orientation adjustment within the store; however, it was only tested with a limited number of product classes, and GPS signals are inherently attenuated in indoor environments, resulting in degraded localization accuracy. Similarly, the system described in [25] used computer vision to read text from a region of interest indicated by the user's hand on a product package. While effective at close range, this approach becomes unreliable at greater distances where model performance degrades, and users may have to touch many products until they touch the target product. The system proposed in [26] utilizes large language models to enable pBLV to discover and learn about products in real-time through tactile and conversational interactions. However, this system requires users to be in close physical contact with products and does not support product detection or spatial guidance from a distance.

In this paper, we present a multimodal wearable assistive system that integrates an object detection model with a VLM to identify products on shelves and help users locate and reach them. The system processes an RGB camera stream and provides both spatialized sonification and spoken feedback to convey the target's relative position. To deliver this functionality, the system operates through a three-phase workflow. In the product searching phase, the system identifies the target product on the shelf and maintains continuous tracking. During the product navigation phase, it provides auditory and verbal cues describing the target's relative position, enabling users to move their hand toward the product. In the product correction phase, the system determines whether the touched item matches the intended product and, if not, offers concise feedback to guide the user toward the correct one. We hypothesize that (1) the integration of YOLO-World detection with embedding similarity matching will achieve high product identification accuracy at typical shopping distances, (2) VLM-based spatial guidance will be able to provide accurate directional feedback, and (3) the correction module will reliably distinguish between correct and incorrect product selections to prevent retrieval errors. To assess the system's performance, we conducted systematic technical evaluations of each module.

## Methods

### Equipment

The system was deployed on an NVIDIA Jetson Orin NX, powered by a Krisdonia 25000mAh power bank for portable operation. Both the Jetson and battery were placed in a backpack for mobility. A USB IMX258 camera was mounted on the shoulder strap of a backpack and was facing forward capturing the scene at 1280×720 resolution. Audio feedback was delivered through SHOKZ OpenComm2 bone-conduction headphones, allowing users to hear guidance while maintaining awareness of their surroundings [10][12][21][27].

**System design**

*Shopping list creation*

To enable the system to identify specific items the user wishes to purchase, a shopping list was created first. This process acquired a sequence of target product images based on user input through a voice-based interface. For speech synthesis and recognition, the system employed two models: KittenTTS for text-to-speech output and OpenAI Whisper [28] for speech-to-text transcription.

*Database.* The system leveraged the Open Food Facts (OFF) database to acquire target product images [29]. OFF is a collaborative, open-access repository containing over four million food product entries from more than 150 countries, with standardized metadata including product names, brands, flavors, packaging types, quantities, and multiple reference images. OFF was selected for its extensive coverage of commonly sold grocery products and its public API that facilitates efficient querying and image retrieval.

*Reference image retrieval.* The system operated on a pre-downloaded JSON catalog derived from the United States subset of the OFF database. For each product request, the user was prompted via spoken instructions to provide the brand name, product name, and optional quantity information. These inputs underwent text normalization consisting of lowercasing, Unicode normalization, punctuation removal, and tokenization. The same normalization procedure was applied to all entries in the OFF catalog.

Following text normalization, candidate products were identified through a filtering pipeline as shown in Figure 1. The brand filtering was applied first, followed by product name filtering and quantity filtering. The quantity filtering was only executed when quantity information was provided by the user. Brand and product name filtering employed fuzzy string matching [30], with similarity thresholds of 0.60 and 0.65 respectively. These thresholds were selected empirically to prevent incorrect product selection while allowing for minor input errors.

Upon completion of the filtering pipeline, the system generated a ranked shortlist of candidates based on aggregate similarity scores. For the top-ranked candidate, visual reference images were retrieved using the product's barcode via the OFF API.

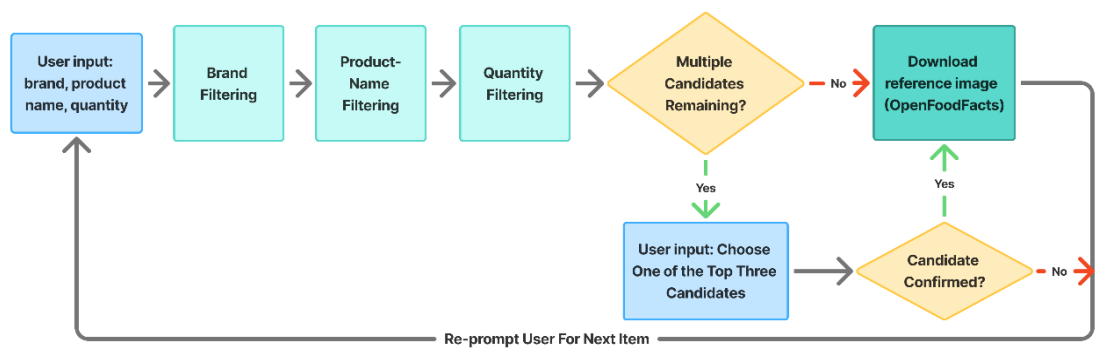

Figure 1. Shopping list creation workflow. Users provide product specifications (brand, name, quantity) via voice input, and the system filters the Open Food Facts database using fuzzy string matching. When multiple candidates remain, users select from the top three matches. Upon successful identification, the corresponding reference image is downloaded for subsequent product matching.

*Product searching*

This section describes how the system identifies target products in users' shopping lists within the store. Product searching consists of product detection, target product matching and product tracking as shown in Figure 2.

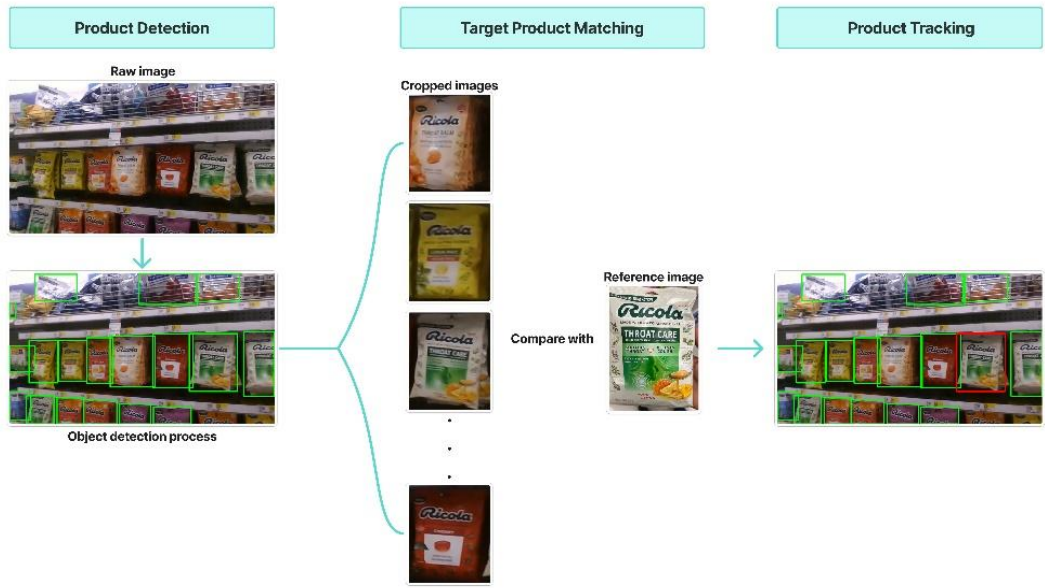

Figure 2. Overview of the three-phase product identification pipeline. Product Detection: YOLO-World detects products on retail shelves. Target Product Matching: Cropped detections are compared against the reference image to find the target product. Product Tracking: The identified target is continuously tracked until new target is set.

*Product detection.* The product detection module was responsible for identifying potential product candidates within the camera feed. We used YOLO-World, an open-vocabulary object detection model capable of detecting arbitrary objects defined through textual prompts [31]. Unlike traditional detectors trained on a fixed set of classes, YOLO-World leverages vision-language alignment to generalize to unseen categories. In our system, the detection prompt was set to include both product and non-product terms, e.g., "product, item, human, body part". The additional non-product classes ("human," "body part") helped the model produce more complete and semantically stable detections in cluttered shelf scenes by avoiding false suppression of hands or other objects partially occluding the product. For each detected object, the model output a bounding box, a class label, and a confidence score; only detections matching product-related classes were retained for subsequent processing.

*Target product matching.* The object matching module compared detected bounding boxes from YOLO-World with reference images downloaded from the shopping list section to determine the most likely product match. This comparison integrated two components: embedding similarity and image color similarity.

For the embedding similarity, we employed a lightweight MobileNetV3-Small encoder to generate compact visual descriptors [32]. The image embedding transformed an image into a numerical feature vector that captures its shape, texture, and global appearance in a high-dimensional space [33]. Images that are visually similar produce embeddings that lie close to one another, enabling efficient retrieval through distance-based comparison. In our system, the classification head of MobileNetV3-Small was removed so that the network output a 576-dimensional feature vector for each image, and cosine similarity was computed between the query embedding that was derived from a cropped detection and each reference embedding in the shopping list.

To incorporate color information as a complementary cue alongside deep learning embeddings, the proposed system used a color histogram matching stage operating in the CIELAB color space [34]. The CIELAB was selected for its perceptual uniformity, wherein Euclidean distances between color coordinates correspond more closely to human-perceived color differences. In this three-dimensional space, the $L^*$ channel encodes luminance while the $a^*$ and $b^*$ channels represent the green-red and blue-yellow chromatic opponent axes respectively.

Rather than computing a single global histogram for each image, the system partitioned both query crops and reference images into three horizontal bands corresponding to the top ($T$), middle ($M$), and bottom ($B$) regions of the product packaging. This spatial partitioning strategy was designed to improve robustness in real-world retail environments, where products may be partially occluded by adjacent items, shelf edges, or the user's hand during retrieval, allowing unobstructed regions to still contribute meaningful color information when other portions are obscured. For each spatial band, a three-dimensional histogram was constructed capturing the joint distribution of all CIELAB channels with $16 \times 10 \times 10 = 1600$ bins for the $L^*$, $a^*$, and $b^*$ channels respectively. The histogram was normalized to unit sum to ensure invariance to region size. For each band $k \in \{T, M, B\}$, the color similarity between the crop histogram $\mathbf{c}^k$ and reference histogram $\mathbf{r}^k$ was calculated using the Bhattacharyya coefficient with equation (1):

$$S_k = \sum_i \sqrt{c_i^k \cdot r_i^k} \tag{1}$$

, where $i$ indexes the histogram bins. This coefficient yields values in the range [0,1], with 1 indicating identical distributions. The composite color similarity score was computed by aggregating the three band scores with equation (2):

$$S_{\text{color}} = w_T \cdot S_T + w_M \cdot S_M + w_B \cdot S_B \tag{2}$$

where the weights $(w_T, w_M, w_B) = (0.3, 0.4, 0.3)$ assign slightly greater emphasis to the central region where distinctive product imagery typically appears. The composite color score was then fused with the embedding similarity to produce the final matching score using equation (3):

$$S_{\text{final}} = \alpha \cdot S_{embed} + (1 - \alpha) \cdot S_{color} \tag{3}$$

Here the $\alpha$ is set as 0.70 to ensure that color serves as a complementary cue rather than the primary discriminator, given that color histograms do not encode shape or texture information.

To optimize processing time, the system ignored detection results with an embedded similarity below 0.50. In addition, both embedding processing and color comparison were GPU-accelerated and batch-processed, enabling real-time inference on Jetson.

*Product tracking*. Once the target product was successfully matched, the system initialized a single-object tracker to maintain a persistent lock on the product across frames [35]. This tracking stage provided temporal stability, and it substantially reduced redundant computation by avoiding continuous re-detection. For object tracking, a Channel and Spatial Reliability Tracker (CSRT) employed a spatial reliability map to handle partially occluded and non-rectangular objects while estimating the object's position frame by frame [36]. Periodic validation through YOLO-World detections occurred every 20 frames to prevent drift and ensure that the tracked region remained aligned with the true product. During re-validation, the system computed a combined embedding and histogram similarity score against the target reference; if this score fell below a threshold of 0.5, or if the tracker had remained stale for more than 80 frames without successful validation, a fresh product detection–matching cycle was triggered to re-anchor the bounding box.

*Product navigation*

Once the target product was identified and its bounding box was stabilized by the tracker, the product navigation module conveyed the product's relative spatial position to help the user orient toward it. This module integrated two complementary auditory channels [12][27]: spatialized sonification and VLM-derived verbal descriptions, both driven by real-time visual and hand-pose signals.

*Sonification feedback.* The sonification feedback provided users with spatial guidance to locate the target product relative to their body position. The system employed a two-stage feedback approach. In the initial stage, the system provided verbal descriptions of the product's location relative to the camera's center, which approximated the user's torso position when the device was placed at chest level. In the subsequent stage, once the user's hand entered the frame, the system transitioned to hand-relative guidance. The system employed MediaPipe Hand, which extracted hand landmarks from the RGB stream [37]. The landmark corresponding to the index fingertip was used as the reference point for evaluating the product's position relative to the user's hand. The system compared the pixel displacement between the index fingertip and the center of the target product's bounding box. This displacement was mapped to stereo-panned beeping tones, where dominance in the left or right audio channel indicated lateral offset. The pitch of the beeping tone increased as the object approached the center of the bounding box of the target object.

*Speech feedback.* To complement these low-level cues, we utilized a VLM to provide short feedback on the target's spatial location within the view of the camera as shown in Figure 3. The VLM integrates a visual encoder with a large language model, enabling it to interpret the camera frame and produce structured natural-language outputs conditioned on a textual prompt [38]. In our system, the prompt partitioned the image into five horizontal and three vertical zones and instructed the model to assign the target to one of these coarse spatial regions (e.g., "far left," "middle," "right," further combined with "upper," "center," or "lower"). To ensure that the VLM produced consistent short-form responses rather than free-form descriptions, we employed prompt engineering strategies including chain-of-thought and few-shot prompting to constrain the model's output space and improve reliability [39]. The resulting text was subsequently converted to speech using the system's text-to-speech module, providing the user with clear spoken feedback such as "upper right".

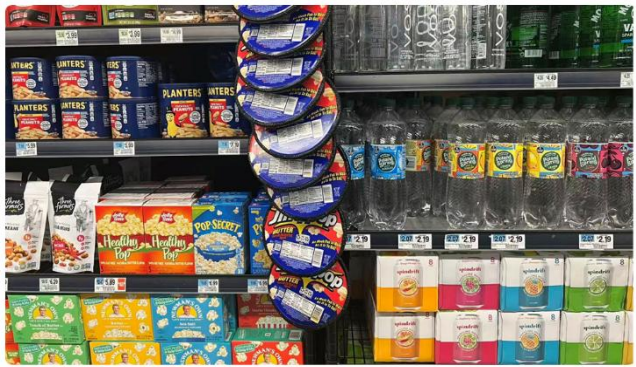

Figure 3. Example of product navigation with VLM. Given the target product (Spindrift Unsweetened Lime Sparkling Water), the VLM analyzes the camera frame and generates a spatial descriptor ("Far right, lower") to guide the user toward the target location.

*Product correction*

The product correction module verified whether the user's hand was interacting with the correct product and provided corrective instructions when necessary. This module operated once the system detected a hand in proximity to a product bounding box. If the index finger's coordinates fell within the target bounding box for three seconds, the system detected that the target product had been touched and informed the user. The system then proceeded to initiate the product detection process for the next item on the shopping list.

If the touch fell on a non-target product, the VLM was triggered to analyze the positional difference between the touched product and the target product in terms of the number of distinct products separating them along the shelf rows and columns, as shown in Figure 4. For small separations (≤4 product types), the feedback used fine-grained phrases such as "Two products to the right". For larger separations (>4 product types), the feedback used coarser descriptors such as "far left" or "far right." Prompt engineering was also applied to this function.

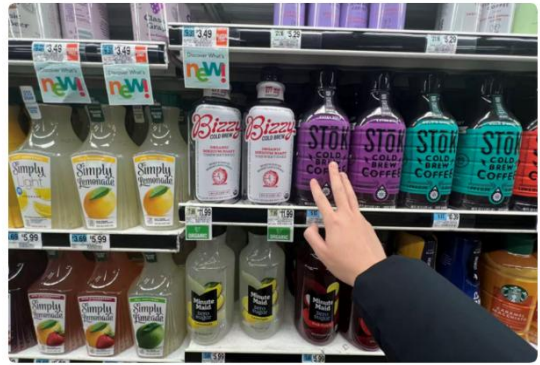

Figure 4. Example of VLM-based product correction. When the user's hand touches an incorrect product (Stok Cold Brew Coffee), the VLM will generate corrective feedback ("Two products to the left, one product down") to guide the user toward the given target product (Simply Lemonade with Strawberry).

**Experiment**

To evaluate the proposed system, we conducted a set of controlled laboratory experiments using a simulated grocery shelf. The experimental setup and evaluation procedure were designed to assess the accuracy and responsiveness of the four major system modules: shopping list creation, product search, product navigation, and product correction.

*Shelf construction*

A physical shelf environment was constructed to approximate common in-store product layouts as shown in Figure 5. The shelf consisted of three tiers, each holding six distinct products placed in small clusters to emulate realistic retail spacing and visual clutter. The shelf measured 1.5 m in width, with each tier spaced 0.4 m in height. The product categories included boxes, bottles, and cans ranging from small to large sizes. Products were positioned at fixed locations, and lighting conditions were constant throughout all trials.

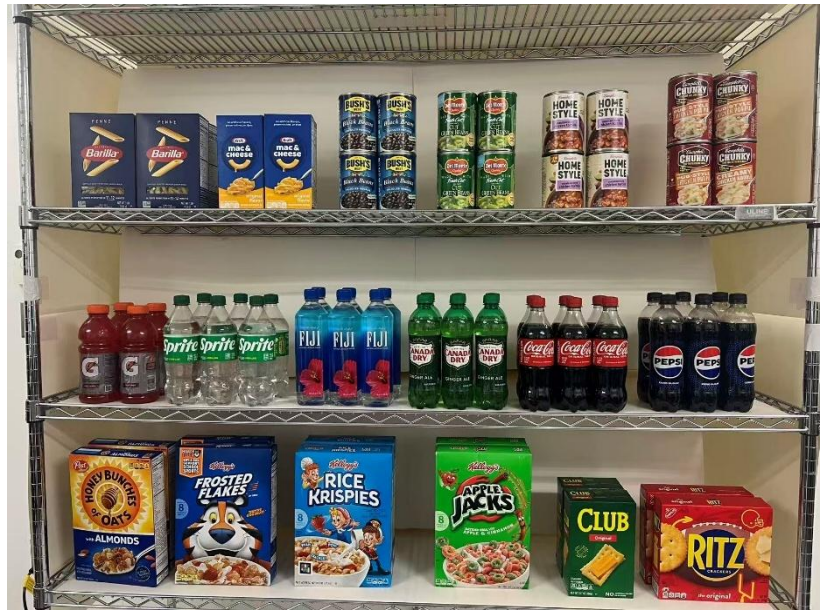

Figure 5. Experimental shelf setup containing 18 distinct product types arranged across three tiers. Products include various packaging formats (boxes, bottles, and cans).

*Validation of shopping list creation*

To evaluate the shopping list creation module, we constructed a test set of 80 products spanning four common grocery categories: Dairy and Eggs, Pantry, Snacks, and Beverages, with 20 products per category. These categories were selected to represent a diverse range of product characteristics, including variations in brand naming conventions, product descriptors, size specifications, and packaging formats commonly encountered in grocery shopping. To minimize selection bias, the test set was generated using a vision-language model prompted to produce representative product examples for each category, rather than being manually curated by the experimenters.

The system was tasked with querying the OFF database. For each query, the system attempted to identify the correct product entry and retrieve its corresponding reference images via the OFF API. A trial was considered successful if the system returned the correct product entry matching the intended brand and product name, with valid reference images available for downstream visual matching. Trials were marked as failures if the system returned an incorrect product, failed to identify any matching candidates, or retrieved a correct product entry lacking usable reference images.

*Validation of product detection module*

The product detection module was evaluated to determine the system's ability to robustly identify all items placed on the shelf across varying user distances and approach angles. We sampled a set of viewing positions forming a conical region in front of the shelf as shown in Figure 6. Specifically, measurements were taken at radius of 0.5 m, 1.0 m and 1.5 m, each at three azimuth angles: 0°, ±30°, and ±60° relative to the shelf normal except for 0.5 m distance. These positions simulated realistic in-store scanning behavior, where a user may not always face the shelf directly. For most viewing positions, the camera was oriented directly toward the center of the shelf, ensuring all 18 products remained within the field of view. However, at the 0.5-meter distance and at the ±60° azimuth angle of the 1.0 m distance, the combination of close proximity and oblique positioning prevented all products from being captured in a single center-facing frame. In these cases, the camera orientation was adjusted to pan across the shelf, ensuring that all 18 products were sequentially brought into the field of view and evaluated.

Prior to evaluation, a shopping list containing all 18 test products was created using the shopping list creation module. For each product on the list, the system queried the OFF database and retrieved corresponding reference images, which were stored locally to serve as visual templates for the product detection module. The evaluation proceeded iteratively, with each of the eighteen products designated as the target in sequence. The accuracy was computed as the proportion of trials in which the module produced correct outputs.

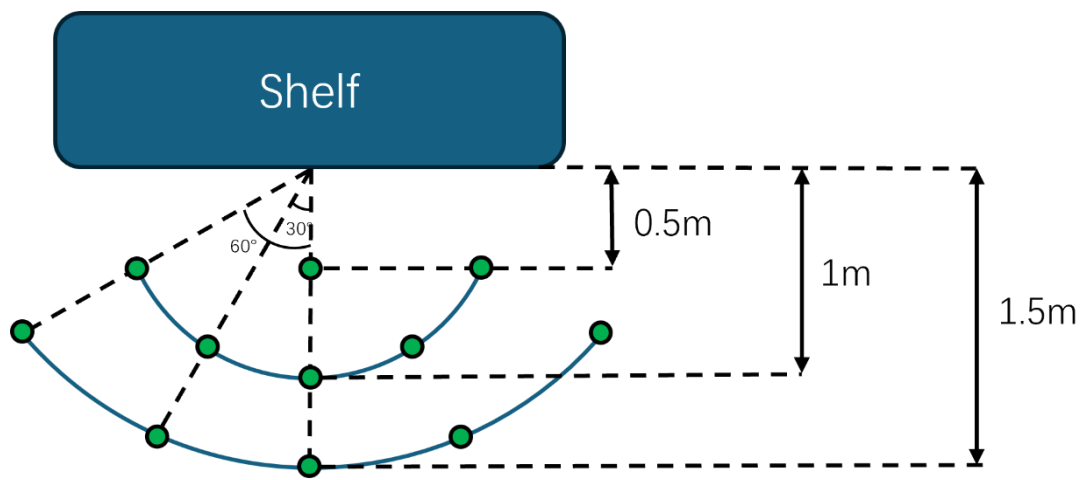

Figure 6. Experimental setup for product detection validation. Green dots indicate camera positions at three distances (0.5 m, 1.0 m, and 1.5 m) from the shelf, with viewing angles of 0°, ±30°, and ±60° relative to the shelf normal. The 0.5 m position was tested only at frontal orientation (0°).

*Validation of object navigation module*

The object navigation module was evaluated to assess the accuracy of the VLM's response which guides users toward target products. The evaluation was conducted at two standing distances of 1.5 m and 1.0 m from the shelf. At each distance, the eighteen products were designated as the target iteratively. For every trial, the system captured a frame containing the target product and generated a verbal spatial descriptor characterizing the target's position within the field of view. Navigation accuracy was computed as the proportion of trials in which the VLM-generated description matched the corresponding ground truth for the target product at the given distance. Five GPT series models (GPT-4o, GPT-4o-mini, GPT-4.1, GPT-5-mini, and o3) were evaluated across both distance conditions to assess comparative performance in the ability to interpret product locations within a cluttered shelf environment and generate accurate directional descriptors relative to the camera's field of view.

*Validation of object correction module*

The object correction module was designed to assess the accuracy of VLM-generated feedback under controlled touch error conditions. The experimenter was positioned 0.5 m from the shelf, representing a typical reaching distance during product retrieval. Due to the constrained field of view of the camera at this proximity, only two shelf tiers could be captured simultaneously within a single frame. Consequently, the evaluation was conducted across two camera configurations: the top configuration, which captured the upper and middle tiers, and the bottom configuration, which captured the middle and lower tiers. The validation comprised two phases, each targeting a distinct error magnitude regime, and five GPT series models were evaluated across both phases.

The first phase evaluated correction accuracy for touch errors occurring within four product-type hops of the target. For each trial, the experimenter designated a target product and subsequently performed an intentional touch on a non-target product. This touch triggered the VLM-based correction function, which generated directional instructions and hop-count estimates. Each of the twelve product categories within a given camera configuration served as a target product in sequence within a given camera configuration, with each target corresponding to seven distinct erroneous touch locations within the four-hop constraint boundary. This yielded a total of 168 trials for the proximate error condition across both the top and bottom configurations. The accuracy considered both directional instruction generation and the correctness of the reported hop count.

The second phase evaluated the correction module under conditions where touch errors exceeded four product-type hops from the target, representing scenarios that triggered coarse directional guidance toward the target's general location rather than precise step-by-step instructions. Target products for this phase were restricted to positions in the leftmost and rightmost columns of the shelf grid to ensure sufficient spatial separation from potential touch locations. Two non-target touch locations were tested per target position, yielding a total of 16 trials across both camera configurations. Evaluation focused on two criteria. First, the system was assessed on whether it correctly identified that the touch error exceeded the four-hop threshold, triggering the coarse feedback mode rather than attempting precise hop-count estimation. Second, the system was assessed on whether the generated directional instruction accurately guided the user toward the target's general region. For example, if the target was located on the far left of the shelf and the user touched a product on the far right, a correct response would indicate that the target is located to the left, such as "the target is far to your left" or "move several products to the left."

**Result**

*Shopping list creation results*

The shopping list creation module was evaluated across four product categories with 20 products each. Table 1 summarizes the accuracy results. The module achieved an overall accuracy of 93.75%, with consistent performance across Dairy & Eggs, Pantry, and Snacks categories at 95.0% each. Beverages exhibited slightly lower accuracy at 90.0%. Failures were primarily attributed to products with ambiguous naming conventions, limited representation in the OpenFoodFacts database, or cases where multiple similar product entries existed without sufficient distinguishing metadata.

| Category | Correct | Total | Accuracy |
|---|---|---|---|
| Dairy & Eggs | 19 | 20 | 95.0% |
| Pantry | 19 | 20 | 95.0% |
| Snacks | 19 | 20 | 95.0% |
| Beverages | 18 | 20 | 90.0% |
| **Overall** | **75** | **80** | **93.75%** |

Table 1. Shopping list creation validation results

*Product detection results*

The product detection results are presented in Table 2. At 0.5 m with a frontal viewing angle (0°), the system achieved perfect detection accuracy at 100%, with no false negatives or false positives recorded. At 1.0 m, accuracy remained at 100% for the frontal position, decreased to 94.4% at 30°, and further declined to 75.0% at 60°, where five false negatives and four false positives were observed. At 1.5 m, accuracy measured 94.4% at 0°, 72.2% at 30° with four false negatives and six false positives. The highest error rates occurred at the maximum distance and angle combination (1.5 m, 60°) with 55.6% accuracy.

| Distance | Angle (°) | Success (count) | Accuracy | FN (count) | FP (count) |
|---|---|---|---|---|---|
| 0.5 m | 0 | 36 | 100% | 0 | 0 |
| 1.0 m | 0 | 36 | 100% | 0 | 0 |
|  | 30 | 34 | 94.4% | 1 | 1 |
|  | 60 | 27 | 75.0% | 5 | 4 |
| 1.5 m | 0 | 34 | 94.4% | 1 | 1 |
|  | 30 | 26 | 72.2% | 4 | 6 |
|  | 60 | 20 | 55.6% | 11 | 5 |

Table 2. Product detection validation results

*Product navigation results*

Table 3 summarizes the accuracy rates for each model at both distance conditions. The evaluation results revealed substantial variation in navigation accuracy across models and operating distances.

At the 1.5-meter distance, the o3 model achieved the highest accuracy at 94.4%, with only one incorrect response out of eighteen trials, followed by GPT-5-mini at 83.3%. The GPT-4o and GPT-4.1 models demonstrated identical performance at 72.2%, while GPT-4o-mini exhibited the lowest accuracy at 38.9%, with eleven incorrect responses. At the reduced distance of 1.0 meter, the o3 model maintained the highest accuracy at 88.9%, with two incorrect responses, followed by GPT-5-mini at 85.6% and GPT-4.1 at 77.8%. The GPT-4o model decreased from 72.2% to 61.1%, representing the second largest performance decline among all models. The accuracy of GPT-5-mini improved from 83.3% to 85.6%, while GPT-4.1 similarly increased from 72.2% to 77.8%. In contrast, GPT-4o-mini exhibited notable degradation, with its accuracy plummeting from 38.9% to 16.7%, producing incorrect responses in 15 out of 18 tests.

| Model | Accuracy 1.5 m | Accuracy 1 m |
| --- | --- | --- |
| GPT-5-mini | 83.3% | 85.6% |
| o3 | 94.4% | 88.9% |
| GPT-4.1 | 72.2% | 77.8% |
| GPT-4o | 72.2% | 61.1% |
| GPT-4o-mini | 38.9% | 16.7% |

Table 3. Accuracy of the VLM output for product navigation

*Product correction results*

Table 4 summarizes the performance of the five tested VLMs across product correction tasks categorized by shelf region and correction complexity. Performance varied notably across models and task conditions.

GPT-5-mini achieved the most consistent results, maintaining accuracy above 86% across all conditions and attaining perfect accuracy in bottom tasks regardless of correction complexity. The o3 model also demonstrated stable performance, achieving 85.7% and 89.3% accuracy for the top conditions and perfect accuracy for bottom corrections requiring fewer than four hops, though its accuracy decreased to 87.5% for larger corrections in the lower shelf region.

Models in the GPT-4 family exhibited considerably greater difficulty with the correction task. GPT-4.1 and GPT-4o produced noticeably lower accuracy, particularly for small-offset corrections in the top region where distinguishing subtle product-type transitions is required, with both models achieving accuracy below 30%. GPT-4o demonstrated inconsistent behavior, achieving 75.0% accuracy for bottom small corrections but dropping to 25.0% for larger corrections in the same region. GPT-4o-mini performed the worst across all conditions, with accuracy ranging from 11.9% to 15.5% in the top region and failing entirely in both bottom conditions with 0% accuracy.

| Model | Top (<4 hops) | Top (>4 hops) | Bottom (<4 hops) | Bottom (>4 hops) |
|---|---|---|---|---|
| GPT-5-mini | 95.2% | 86.9% | 100% | 100% |
| o3 | 85.7% | 89.3% | 100% | 87.5% |
| GPT-4.1 | 27.4% | 46.4% | 62.5% | 62.5% |
| GPT-4o | 26.2% | 42.9% | 75.0% | 25.0% |
| GPT-4o-mini | 15.5% | 11.9% | 0% | 0% |

Table 4. Accuracy of the VLM output for product correction

**Discussion**

The proposed multimodal assistive system was designed to support independent product retrieval for people who are blind or have low vision through three integrated modules: product detection, product navigation, and product correction. The technical evaluation results support all three hypotheses: (1) product detection achieved near 100% accuracy at frontal viewing angles across all tested distances, confirming high identification accuracy; (2) VLM-based navigation achieved up to 94.4% accuracy with the o3 model, demonstrating reliable spatial guidance; and (3) the correction module with GPT-5-mini achieved above 86% accuracy across all conditions, validating its ability to distinguish correct from incorrect product selections.

*Validation results analysis*

The system was technically validated across three core modules. Product detection achieved near perfect accuracy at 0° viewing angles, remaining above 90%. Both false negatives and false positives increased progressively as viewing conditions became more demanding. Product navigation demonstrated strong model-dependent performance, with the o3 model achieving the highest accuracy at 94.4% at 1.5 m and 88.9% at 1.0 meter, while GPT-5-mini maintained consistent performance across both distances. Product correction accuracy revealed similar model stratification, with GPT-5-mini achieving above 86% accuracy across all conditions and near perfect performance for bottom shelf corrections, whereas GPT-4 family models struggled considerably, particularly with small-offset corrections requiring precise product-type differentiation. These validation results demonstrate the system's potential as an assistive tool to help pBLV locate and acquire target shopping items in the store.

**Target *p*roduct *m*atching**

As described in the Method section, the system compares cropped images from object detection bounding boxes against reference images downloaded from the OFF database. Consequently, the quality of these reference images directly influences the similarity computation. Images in the OFF database are contributed voluntarily by users without stringent quality control requirements, resulting in considerable variability. For instance, some reference images include the uploader's hand partially occluding the product, while others contain substantial portions of irrelevant background. Such extraneous visual elements introduce noise into both the embedding module and the color histogram module, potentially degrading matching accuracy.

Furthermore, when the system's camera is positioned far from the shelf or captures the scene at an oblique angle, the resulting frame contains fewer pixels representing the product. This reduced resolution hampers the object detection model's ability to extract discriminative features, leading to missed detections [40]. Even when detection succeeds under these conditions, the limited pixel count in the cropped region compromises subsequent similarity matching, ultimately resulting in failed target identification.

**VLM-based spatial reasoning for assistive applications**

The strength of VLMs in relative spatial reasoning has broader implications for assistive shopping technology while there is a research gap using VLMs for the complete shopping journey.

Beyond product localization and retrieval, when users are at the store entrance, VLMs could describe the general layout of the retail environment, identifying the relative positions of different sections such as "produce is to your left, dairy is straight ahead, and checkout is to your right." When navigating aisles, VLMs could provide contextual awareness by describing nearby shoppers, cart positions, or temporary obstacles, enabling pBLV to navigate crowded spaces more confidently. During product selection, VLMs could describe the relative positions of different sizes or variants of the same product, such as "the family size is one shelf above the regular size" or "the low-sodium version is two products to the left." VLMs could also assist with price comparison by describing the spatial relationship between products and their corresponding price tags or help users identify promotional displays and sale items relative to their current position. At checkout, VLMs could guide users toward available registers or self-checkout stations by describing queue lengths and relative positions, such as "the shortest line is three stations to your right." VLMs could also support payment interactions by describing the relative positions of card readers, PIN pads, and receipt dispensers, reducing reliance on store staff assistance.

However, the VLM component also exhibits notable limitations in our validation tests. Despite their remarkable capabilities in general visual understanding tasks, current VLMs demonstrate systematic deficiencies in spatial reasoning [41]. Research has shown that some state-of-the-art VLMs struggle with basic spatial relationship inference, achieving accuracy only marginally above random chance on perspective-dependent spatial localization tasks [42]. This limitation is particularly pronounced in our application context, where the model must accurately determine which horizontal and vertical zone contains the target product.

In addition, VLMs also failed to count the exact number of individual items between the incorrectly touched product and the target product on the shelf correctly, which would have provided users with more precise corrective guidance. However, all tested VLMs failed to achieve acceptable accuracy on this task based on our internal test which is consistent with prior findings that the VLMs exhibited systematic counting failures, with accuracy declining markedly as the number of intervening items increased [43].

*Form factors for shopping assistive system*

Form factors are a primary consideration when determining the camera's position in our system. In this study, the camera was mounted on the backpack's strap near the chest position to capture the product due to its stable field of view and alignment with the intended travel path and direction of motion of the user [44]. However, such configuration could introduce potential occlusion to the product when moving arms or hands enter the camera's field of view during product manipulation.

Our system can also be integrated with a head-mounted camera or smart glasses such as Meta Glasses would liberate both hands for product manipulation while aligning the camera direction with the user's attentional focus [45]. However, head-mounted configurations require that the user's head orientation remains relatively aligned with the target, necessitating deliberate head movements to locate objects outside the camera's field of view [46]. Furthermore, rapid head rotation can induce motion blur in captured images and compromise recognition accuracy for tasks requiring high-frequency visual details [47].

Handheld smartphones represent another promising solution, with applications such as Be My Eyes [48], and Microsoft's Seeing AI [49] demonstrating considerable utility for daily tasks including object finding. However, handheld smartphone-based solutions present inherent challenges for pBLV users [50]. Handheld operation requires users to dedicate one hand to holding the device, which limits their ability to simultaneously interact with products or use other mobility aids such as white canes. Therefore, this setup may not be user-friendly for pBLV navigating stores, as it requires them to ensure their own safety while also verifying product information. Furthermore, for systems such as ours that rely on tracking the relative position of the user's hand in relation to the target product, this core functionality would be compromised in handheld configurations. The camera must simultaneously maintain a clear view of the reaching hand while capturing the target product, a dual requirement that a single handheld device cannot reliably satisfy. One potential adaptation could leverage changes in the smartphone's position as a proxy for the user's movement toward the target, though this approach remains an area for future investigation.

**Integration with multi-scale navigation ecosystems**

The proposed system can work as a standalone solution but can also be integrated with a hierarchical, multi-scale assistive navigation ecosystem. The navigation requirements of pBLV span multiple spatial granularities, from macro-scale urban wayfinding to micro-scale object localization. Effective deployment of the proposed shelf-level product identification system would therefore necessitate seamless integration with complementary technologies addressing the preceding phases of the shopping trip.

For outdoor navigation from residence to retail destination, the system could interface with existing audio platforms such as BlindSquare [51], or Lazarillo [52], which employ three-dimensional audio beacons and environmental callouts to support ambient awareness and destination-directed wayfinding. Upon entering the retail environment, where GPS signals attenuate below usable accuracy thresholds, indoor navigation systems would assume navigational responsibility, guiding users through store aisles to locate the correct shelf containing their desired products [53][54]. The proposed system then engages at the final and most granular spatial scale, addressing the last-meter problem wherein users must transition from aisle-level orientation to precise product-level localization and retrieval [55]. Such hierarchical decomposition would enable each component to be optimized for its respective spatial resolution and sensing modality while collectively providing continuous navigational support across the complete shopping workflow.

**Conclusion**

This paper presented a multimodal wearable assistive system designed to help pBLV independently locate and retrieve products in physical stores. The system integrates object detection with VLMs to enable three-phase product assistance: product search, product navigation, and product-targeting correction. Technical evaluation demonstrated promising performance, with product detection achieving near-perfect accuracy at close range and maintaining high accuracy when facing the shelves at 0° viewing angles within 1.5 m. These results demonstrate the system's potential to address precise product-level localization and retrieval, though several limitations remain. Future work will focus on conducting user studies with pBLV participants in authentic retail environments, improving matching algorithms, and integrating the system with multi-scale navigation ecosystems to enable seamless end-to-end shopping assistance.